# A High-Flux and High-Efficiency Setup for Magneto-Infrared Spectroscopy

Zeping Shi[1†], Wenbin Wu[1†], Zhiwei Zhang[1†], Yuhan Du[1†], Chenyao Xu[1], Guangyi Wang[1], Mingsen Zhou[1], Congming Hao[1], Xianghao Meng[1], Xiangyu Jiang[1], Chunhui Pan[2], Wei Lu[2], Hao Shen[1], Haifeng Pan[1], Zhenrong Sun[1], Junhao Chu[3,4], Xiang Yuan[1,3*]

[1] State Key Laboratory of Precision Spectroscopy, East China Normal University, Shanghai 200241, China

[2] Multifunctional Platform for Innovation Precision Machining Center, East China Normal University, 200241, Shanghai, China

[3] School of Physics and Electronic Science, Key Laboratory of Polar Materials and Devices, Ministry of Education, East China Normal University, Shanghai 200241, China

[4] Institute of Optoelectronics, Fudan University, 200438, Shanghai, China

[†]These authors contributed equally to this work.

[*]Correspondence and requests for materials should be addressed to X. Y. (E-mail: xyuan@lps.ecnu.edu.cn)

**Abstract**

We report the design and implementation of a high-flux, high-efficiency magneto-infrared spectroscopy system optimized for broadband measurements in high magnetic fields. The setup integrates a Fourier transform infrared spectrometer, a 12 T cryogen-free superconducting magnet, precision-polished and gold-plated light tubes, custom-designed reflective focusing modules for Faraday and Voigt geometries, and an external multi-detector chamber with motorized selection. Optical throughput is maximized by reducing light tube loss from 65.5% m$^{-1}$ to 22.0% m$^{-1}$ via abrasive flow and mechanical polishing followed by gold electroplating, and by adopting a single-on-axis parabolic-mirror Faraday module that increases the effective numerical aperture from 0.14 to 0.36, enhancing collection efficiency by nearly an order of magnitude. An eight-position motorized sample stage and fully automated control over magnetic field, temperature, optical path, and detector choice enable high-throughput measurements without repeated warm-ups. The optimized configuration achieves a root-mean-square noise level of 0.0061% in a 2-minute integration for a 40% reflectivity sample, corresponding to a signal-to-noise ratio exceeding $1.6 \times 10^4$. System capabilities are demonstrated by resolving weak replica bands in EuCd$_2$As$_2$ and faint Landau level transitions in LaAlSi.

## I. INTRODUCTION

Infrared spectroscopy is an indispensable tool in modern experimental science[1,2]. Across physics, chemistry, and materials science, it provides direct access to vibrational, electronic, and collective excitations over an energy range that bridges the gap between microwave and visible light[3–9]. In condensed matter physics, infrared measurements probe a wide variety of quasiparticle dynamics, including phonons, excitons, interband transitions, and collective electronic modes[10–15], thereby offering crucial insights into the electronic structure and low-energy interactions of solids.

When combined with a magnetic field, infrared spectroscopy becomes an even more powerful probe of matter[16–22]. Magneto-infrared spectroscopy can access fundamental phenomena[23–26] such as electron paramagnetic resonance (EPR), cyclotron resonance (CR), and Landau quantization, enabling the determination of carrier effective mass, Fermi velocity, and g-factors, as well as the study of symmetry breaking and topological phase transitions. It can also probe chiral optical responses in diverse magneto-optical materials, study vibrational dynamics of reactive radicals, and reveal biological nanosystem signatures linked to evolution and disease, while also offering insights into magnetized plasmas and interstellar media in astronomy and planetary science. High magnetic fields can be provided by large national facilities[15,27–32], such as the European Magnetic Field Laboratory in Grenoble and the National High Magnetic Field Laboratory in Tallahassee, or by smaller laboratory-scale platforms[33,34] based on superconducting magnets. The latter are essential for routine experiments and long-term studies.

A persistent limitation in magneto-infrared spectroscopy is the degradation of signal-to-noise ratio in high magnetic fields. In conventional magneto-optical setups, an Fourier transform infrared spectrometer (FTIR) coupled to a magnet with optical windows allows ex-situ beam delivery[35–37], but the achievable field is limited. For high-field solenoid magnets, light must traverse long, narrow light tubes to reach the sample space, suffering substantial losses from multiple reflections. The FTIR source, typically a blackbody emitter at a fixed temperature, offers limited scope for increasing brightness. Moreover, the magnet bore is extremely confined, restricting the use of large optics and making it difficult to achieve efficient focusing. This constraint also increases the likelihood of collecting unwanted stray light. In addition, infrared detectors generally have small active areas, and conventional focusing optics in magneto-infrared setups have low effective numerical aperture, which limits detection efficiency. These issues are particularly severe in the mid-infrared regime.

Another major challenge is measurement-efficiency. High-field magneto-infrared experiments are time-consuming, involving not only field sweeps and spectral acquisition, but also repeated warming and cooling for sample changes. The advent of closed-cycle, cryogen-free magnets has enabled long-duration, unattended operation, but most existing systems still accommodate only a few samples per probe, lack automated sample exchange, and cannot switch detectors without manual intervention. As a result, the measurement throughput remains low.

Here we report the development of a high-flux, high-efficiency magneto-infrared spectroscopy system designed to address these limitations. The system consists of an FTIR spectrometer, a superconducting solenoid magnet, light tubes, custom-designed reflective focusing modules for Faraday and Voigt geometries, and an external multi-detector chamber. To maximize optical throughput, we applied abrasive flow polishing and mechanical polishing to the light tube interior, followed by gold electroplating, reducing per-meter loss from 65.5% to 22.0%. A single-parabolic-mirror focusing design in the focusing modules increases the effective numerical aperture to 0.36 and boosts optical collection efficiency by nearly an order of magnitude relative to conventional designs. A motorized, ring-shaped sample stage accommodates 8 samples per probe and enables automated switching without warming the system. The external detector chamber allows large effective numerical aperture collection and in-situ detector changes via a motorized translation stage. The entire setup is controlled by a LabVIEW-based program capable of fully automated measurements over a user-defined range of magnetic field, temperature, optical configuration, sample selection, and detector choice. These innovations significantly enhance both the signal-to-noise ratio, particularly in the mid-infrared range, and the measurement efficiency, enabling the detection of field-induced optical features with amplitudes on the order of $10^{-3}$. Under optimized conditions, a minimum root-mean-square noise level of 0.0061% was achieved for a moderate-reflectivity sample within a 2-minute integration. It should be emphasized that in the optimized configuration, only one temperature ramp is required for the measurement of 8 samples, thus greatly improving the overall efficiency. Thus, the average measurement time per sample was reduced to one-fourth of its previous value. The system performance is further demonstrated by resolving weak replica bands (0.15% amplitude) in $EuCd_2As_2$ and faint Landau level transitions (0.06% amplitude) in LaAlSi.

**II. EXPERIMENTAL SETUP**

The magneto-infrared spectroscopy system integrates five primary subsystems, including an FTIR spectrometer, a cryogen-free superconducting solenoid magnet, light tubes for optical transmission, custom-designed reflective focusing modules for Faraday and Voigt geometries (either Faraday or Voigt geometries), and an external multi-detector chamber (Figs. 1 and 2).

The spectrometer employs a Michelson interferometer with a silicon carbide infrared source (Globar) or tungsten filament lamp, a beam splitter, a fixed mirror (M4), and a movable mirror (M5). The modulated infrared beam is directed through an optical periscope assembly (mirrors M6 and M7) to the top of the superconducting magnet. The magnet is a cryogen-free, 12 T superconducting solenoid magnet equipped with a variable temperature insert (VTI), enabling continuous temperature control of the sample from 2 K to 300 K.

Incident light enters the magnet vertically through an off-axis parabolic mirror (M8) and a diamond window that isolates the magnet vacuum from the incident optical path. The vertical optical transmission is defined by a 1340.5 mm long light tube with an inner diameter of 19 mm. Light reaches the focusing module at the magnet center either directly or after a small number of reflections from the light tube inner wall. Depending on the experimental geometry, dedicated focusing modules for Faraday or Voigt configurations direct the light onto the sample and collect the reflected beam. The collected light is further guided through the exit light tube, redirected by a mirror (M9), and sent toward the external detector chamber.

Before entering the detector chamber, the light passes through a second periscope assembly (mirrors M10 and M11), which delivers the beam to an off-axis parabolic mirror (M12) mounted on a motorized translation stage. The detector chamber houses multiple infrared detectors, including a liquid-nitrogen-cooled mercury-cadmium-telluride (MCT) detector, a liquid-nitrogen-cooled InSb detector, and a room-temperature Si diode detector. The motorized stage positions M12 to focus the incoming beam onto the active area of the selected detector.

The entire optical path outside the magnet is maintained at a pressure of approximately 300 Pa by a mechanical pump to minimize absorption from water vapor and $CO_2$. The section inside the VTI is protected by helium exchange gas. The interferogram is recorded by synchronizing the M5 mirror position with the detector signal, and the resultant spectrum is obtained via Fourier transformation.

To achieve high signal-to-noise ratio, particularly in the mid-infrared range, the system is optimized to maximize optical throughput. Specific techniques target the intrinsic limitations of magneto-infrared systems, namely, weak optical flux in high-field magnets, substantial transmission losses in long light tubes, significant stray-light contamination, reduced focusing efficiency in the constrained sample space, finite numerical aperture at the detector, and extra idle times during probe exchange. Our design incorporates improvements in the light transmission assembly, focusing modules, detector chamber, and software control to overcome these limitations.

## III. SETUP DESIGN AND COMPONENTS
### A. Light tube polishing and gold coating
The light tubes serve as the primary conduits for delivering infrared beam into and out of the bore of the magnet. In high-field solenoid magnets, the diameter of bore is typically limited to ~50 mm, imposing severe constraints on optical access. To accommodate both incident and exit paths within this finite space, each light tube was designed with an inner diameter of 19 mm and a wall thickness of 0.5 mm, fabricated from H62 brass for its non-magnetic properties and ease of machining. However, the combination of long propagation length (1.34 m) and narrow bore inevitably results in substantial optical losses due to multiple reflections from the inner wall of the light tubes. Untreated brass exhibits only moderate infrared reflectivity, and its geometry makes high-quality polishing and coating technically challenging.

To overcome these limitations, we implemented a two-stage polishing procedure followed by electrochemical gold plating to produce a high-reflectivity, low-loss light tube. Abrasive flow polishing was first employed to force a semi-solid, abrasive-laden medium through the full 1.34 m tube length, removing axial machining marks and gross irregularities. This was followed by mechanical polishing to smooth fine transverse imperfections using a custom mandrel, a wool cylinder sleeved over a hardened steel rod, charged sequentially with diamond abrasives from 500 to 10,000 grit. The tube was rotated on a precision lathe while the mandrel was reciprocated axially, and infrared thermal imaging was used to verify tight contact, indicated by locally increased temperatures of ~40 °C. After cleaning and drying, the tubes were chemically degreased, acid-pickled, and activated, and a nickel underlayer was applied to block zinc diffusion. Gold electroplating in a gold cyanide bath, with optimized current density and plating time, produced a dense, uniform, high-purity gold coating. Post-plating, the tubes were rinsed in deionized water, chemically passivated, and dried under hot air atmosphere or vacuum.

This process reduced optical loss from 65.5% m$^{-1}$ for the untreated brass tubes to 22.0% m$^{-1}$ for the polished and gold-plated tubes, yielding a nearly threefold improvement in throughput per meter, markedly enhancing infrared transmission. The loss coefficients were determined by directly measuring the transmitted optical power using a calibrated optical power meter positioned at the exit of each tube. During the measurement, a collimated mid-infrared beam from the FTIR source was directed into the test tube, and the transmitted power was compared with incidence. Over the full 1.34 m path length in the actual system, this improvement more than doubles the delivered optical power, which is critical for enhancing the signal-to-noise ratio where source brightness is inherently limited.

**B. Single-on-axis-parabolic-mirror Faraday focusing module**

In the Faraday configuration, the focusing module directs the incident beam onto the sample and collects the reflected beam for transmission through the exit light tube. In superconducting magnets, the available sample space is extremely limited because the magnetic field strength scales inversely with coil size. This restriction makes efficient focusing challenging and can cause substantial loss of optical power for millimeter-scale or smaller samples, thereby degrading the signal-to-noise ratio. Furthermore, conventional focusing modules are inevitable to collecting stray light scattered within the confined bore, which can mix with the desired signal and obscure weak spectral features.

To address these limitations, we designed a compact optical system for Faraday geometry (Fig. 3), incorporating a single on-axis parabolic mirror for focusing and re-collimation using opposite mirror halves. Reflector locking component, as illustrated in figure, is designed to secure and lock the parabolic mirror in place. Prior to each experiment, fine adjustment of the mirror tube along the vertical direction is necessary to ensure that the infrared beam is focused on the sample surface. To prevent any shift of the parabolic mirror and the tube during the experiment, the nut on the reflector locking component is threaded in the opposite direction to that of the mirror tube. Tightening the nut provides a stop and firmly locks the mirror in place. In this configuration, the incoming light from the incident light tube passes through the central aperture of the ring-shaped sample stage and is directed to the on-axis parabolic mirror mounted at the base of the module. One half of the mirror focuses the incoming beam onto the sample, while the other half collimates the light reflected from the sample, enabling efficient bidirectional beam manipulation within a minimal optical path length. The parabolic mirror, coated with a 500 nm-thick gold layer, has a diameter of 45 mm, a focal length f = 24 mm, and an effective numerical aperture of 0.36. For an FTIR aperture of 6 mm, the focused spot diameter on the sample is approximately 2.5 mm, significantly increasing the optical power density at the sample surface. A light baffle positioned above the mirror prevents stray light with large incident angles, thereby preventing them from coupling into the exit light tube without interacting with the sample.

The ring-shaped sample stage accommodates up to eight samples (each up to 3 × 3 mm²) and is rotated by an external drive system. The motor outside the magnet rotates the drive rod, whose lower end is tightly coupled to the top of a vertical leadscrew. The rotation of the drive rod drives the leadscrew, which in turn rotates the transmission gear mounted at its lower end. This gear meshes with the ring gear that is fixed to the annular sample stage, thereby enabling horizontal rotation of the sample stage within the confined magnet bore. Due to the extremely limited room inside the magnet bore, the ring gear is deliberately offset from the incident light waveguide rather than being perfectly coaxial, enabling the installation of the gear mechanism. The stage is mounted on low-friction PEEK bearings, with its axial position stabilized by titanium alloy spring clips (TC4) to maintain the sample surface precisely at the parabolic mirror focus. The stage assembly can be quickly detached from the gear set for rapid sample exchange. To address potential reflections from the sample holder during experiments, the non-sample regions were covered with carbon adhesive tape, which significantly suppressed spurious reflections.

Optical simulations were performed to compare our design with a conventional Faraday reflection module (Fig. 4a, b). In the simulations, the incident beam was modeled with an 8 mm diameter, uniform spectral intensity corresponding to a blackbody source, and wavelength of 10 μm (1,000 cm$^{-1}$). The divergence half-angle was set to R ≈ 3° and varying the R does not qualitatively affect the comparison results. Ideal mirror surfaces (IRGold, reflectivity 98.99%) and a perfectly reflecting sample were assumed, without accounting for multiple scattering between optical components. In the conventional design, significant beam clipping and scattering occurred at the inclined light tube section, leading to a large spot size on the sample and inefficient collection of the reflected beam. In contrast, the single-on-axis-parabolic-mirror design ensured that the majority of the incident light illuminated the sample and that most reflected light was coupled efficiently into the exit light tube.

Simulation results show that the optical collection efficiency, the ratio of power entering the exit path to the incident power, increased from 8.66% for the conventional design to 75.67% for our module, representing nearly an order-of-magnitude improvement. The power density profile at the exit tube cross-section exhibits a more circular and symmetric spot with significantly higher peak intensity (Fig. 4c, d), indicating improved beam quality and effective suppression of beam distortion. By improving both the focusing quality and collection efficiency, this module significantly enhances the optical throughput and signal-noise ratio, particularly in the mid-infrared regime.

**C. Off-axis-parabolic-mirror Voigt focusing Module**
In certain experiments, it is necessary to perform magneto-infrared measurements in the Voigt geometry, where the light propagation direction is perpendicular to the magnetic field. The Voigt focusing module (Fig. 5) directs the incident beam from the light tube onto the sample surface via an off-axis parabolic mirror (OAPM) with a 90° off-axis angle, 20 mm diameter, and 20 mm focal length. The reflected light from the sample is collected by a second, symmetrically positioned OAPM and collimated into the exit light tube. The two parabolic mirrors share a common focal point in space. This basic optical layout was originally designed by Mykhaylo Ozerov[38]. As in the Faraday module, a graphite-coated light baffle is also incorporated to prevent stray light from bypassing the sample.

The plan view in the dashed box of Fig. 5a illustrates the beam path more clearly. Unlike the Faraday module, where a gear set enables horizontal rotation of the sample stage, the Voigt module adopts a different driving mechanism. Here, the internal thread of the drive rod engages directly with the leadscrew. When the drive rod is rotated, it translates the leadscrew vertically, causing the sample stage to move up and down and each sample sequentially passes through the common focal point of the two OAPMs. Therefore, the sample switching can be realized by a single drive rod for both Faraday and Voigt configurations.

The sample holder features inclined side surfaces adjacent to the mounting plane. These slanted surfaces redirect most of the incident light that does not strike the sample, thereby preventing spurious reflections from contaminating the collected signal. Additionally, similar to the Faraday module, carbon adhesive tape is employed to suppress unwanted reflections from the focusing screw and surrounding surfaces. The rod incorporates an M3 × 0.5 internal thread with a length of 6.6 mm. Rotation of the rod drives the leadscrew and sample holder upward or downward. When the leadscrew and holder approach the upper thread limit, the top of the screw contacts an elastic upper limit mechanism containing a phosphor-bronze spring clip, which pushes the screw downward to ensure that the male thread never disengages from the female thread in the rod. Similarly, when approaching the lower thread limit, the dovetail slider at the bottom of the sample holder contacts a 0.3 mm thick TC4 titanium alloy spring plate (with adjustable deflection) in the lower limit mechanism, pushing the sample holder upward to prevent disengagement of the threads.

To compensate for variations in sample thickness, which could otherwise prevent all samples from lying at the mirror focal plane, the sample holder includes an additional focusing screw. Samples are mounted on the holder end face with GE-varnish, and the screw is rotated to align the sample surface precisely with the shared focal point of the two OAPMs. The OAPM mounts feature special openings that allow fine angular adjustment of the mirrors. To improve positional stability of the sample holder during motion, two phosphor-bronze spring clips are added between the dovetail slider and the dovetail groove, providing bidirectional preload to eliminate clearance while maintaining smooth motion.

**D. External detector configuration**

As shown in Fig. 6, two common optical layouts in existing magneto-infrared setups are the beam-splitting external-detecting configuration and reflective internal-detecting configuration. In the beam-splitting external-detecting design, the incident light passes through a beamsplitter before reaching the sample, and the reflected light from the sample is directed to the detector via reflection from the beamsplitter. This arrangement inevitably loses three-quarters of the optical flux due to the combination of transmission and reflection losses at the beamsplitter.

In the reflective internal-detecting configuration, the light interacts with the sample inside the magnet and the reflected beam is collected directly by a detector placed within the magnet bore, thereby benefiting from a shorter optical path[39]. However, because the detector is mounted inside the magnet, this approach only allows the use of a single detector at a fixed operating temperature. This precludes variable-temperature spectroscopy on the sample and limits the ability to switch to detectors optimized for other spectral ranges. Furthermore, detectors placed inside the magnet must be carefully protected against temperature changes and mechanical vibrations, requiring cautious cooldown procedures as well as relatively complex vibration isolation. For example, in Landau quantization studies of topological materials, the sample often needs to be at cryogenic temperatures near 4 K, while the quantization features of quasiparticles with small effective mass, large Fermi velocity, or a finite energy gap often locates in the mid-infrared regime. Many detectors optimized for the mid-infrared require operation at liquid-nitrogen temperature, making the internal-detector approach impractical in such cases (Fig. 6a and Fig. 6b).

In our system, we adopt a reflective external-detecting configuration (Fig. 6c). Light enters and exits the magnet via the light tubes, and the outgoing beam carrying the sample information is focused by an additional OAPM onto an external detector. This design combines the advantages of high optical flux (avoiding beamsplitter losses), high collection efficiency at the detector, and compatibility with both variable-temperature measurements and multi-band magneto-infrared spectroscopy, as detectors can be selected according to the spectral range of interest. It also provides valuable flexibility for the design and optimization of the focusing modules, as it avoids occupying the extremely limited space inside the bore of magnet.

Halving the radius of the light tube would, in principle, reduce the transmitted flux to one-quarter if the optical power density were constant and wall reflections were neglected. In our system, two factors suppress this reduction. First, the incident beam is pre-focused by parabolic mirror M8 such that the beam spot entering the tube is already smaller than the tube diameter; therefore, doubling the tube diameter would not appreciably increase the captured flux. Second, the light tube inner wall was treated by abrasive-flow and mechanical polishing followed by gold electroplating, which strongly reduces reflective losses. As a result, the increased number of reflections in a narrower tube does not lead to substantial additional attenuation. Furthermore, compared to the beamsplitter configuration, our reflective design offers the added advantage that it is not restricted to a limited spectral window, since infrared beamsplitter typically operate effectively only within specific frequency ranges, whereas our configuration can, in principle, cover the entire infrared regime.

**E. Detector chamber**

Even with optimized optical throughput in the focusing and transmission stages, the goal remains to maximize the detection signal-to-noise ratio, making the detector configuration critical. In both internal and external detector layouts, magneto-infrared systems often face the limitation of a small effective numerical aperture for detectors. Since the active area of most infrared detectors is typically only 1-2 $mm^2$, the collection efficiency is inherently low.

To address this issue, we developed an external detector chamber (Fig. 7). The quasi-collimated beam exiting the 30 mm-inner-diameter light tube, with a finite divergence angle, first reflects from a large flat mirror (50 × 75 mm) and enters the detector chamber. It is then directed onto an OAPM with a focal length of 45 mm. The OAPM is mounted in a vacuum-compatible epoxy resin holder attached to a motorized linear translation stage, allowing it to focus the beam onto the active area at the center of the selected detector. The translation stage enables automated switching between a MCT detector, a InSb detector, and a Si detector, all of which are mounted on a high-precision three-axis positioning stage.

The high repeatability of the translation stage in vacuum ensures precise and stable alignment between the detectors and the light tube, improving long-term system stability. This design eliminates positional shifts that would otherwise occur when switching detectors manually, which is especially important for collecting comparable spectra of the same sample under different experimental conditions. Our tests confirm that the system exhibits no noticeable drift associated with changes in vacuum level.

**F. Automated sample and experimental parameter switching**

In magneto-optical spectroscopy, measurements are typically performed by acquiring Fourier-transform spectra at different magnetic fields and temperatures. When such measurements are controlled programmatically, experimental efficiency can be significantly improved. Beyond the time required for spectral acquisition and magnetic field sweeps, each sample exchange in a conventional system requires removing and reinserting the sample probe, which in turn necessitates warming and cooling the VTI to prevent condensation or ice formation on critical components. Consequently, the ability to integrate as many samples as possible on a single probe and to switch between them automatically without warming the system can greatly increase measurement efficiency.

To this end, our system implements program-controlled switching of multiple parameters: magnetic field (-12 to 12 T), temperature (2-300 K), sample position (#1-#8), beamsplitter (Si/KBr/CaF$_2$), light source (globar/tungsten), and detector (MCT/InSb/Si diode). This approach fully exploits the advantages of a closed-cycle magnet system, enabling unattended multi-sample, multi-band measurements under varied environmental conditions.

In both Faraday and Voigt focusing modules, a motorized eight-position sample holder is incorporated. In the Faraday module, the drive rod is rotated under program control to deliver torque via the drive rod to the gear set, which in turn engages with the ring-shaped sample stage. As shown in Fig. 8a and Fig. 8b, the ring is a hollow circular structure with a central opening to allow the incident beam to pass through. Samples are mounted on the outer ring surface, whose edge incorporates evenly spaced straight gear teeth that mesh with the gear set. The ring can be detached for convenient sample loading. Eight sample positions are evenly distributed along the 3 mm-wide ring surface, and samples are fixed with GE-varnish so that their surface normal point directly downward. As the ring rotates, each sample is sequentially positioned at the focal point of the parabolic mirror for measurement.

In the Voigt module (Fig. 8c), the drive rod is translated vertically under program control to move each sample into the shared focal point of the two OAPMs. Compared to previous designs, the threaded section of the drive rod has been shortened, and the sample holder's drive rod employs a half-threaded design. This ensures that, upon reaching the mechanical limit, further rod rotation disengages the threads rather than locking or even breaking the screw, thereby improving system safety. The multi-seat sample mount increases the number of compatible samples, allowing up to eight samples to be measured in a single cooldown. Furthermore, the sample mount is separable from the sample holder, enabling preparation or replacement of samples without removing the entire holder from the focusing module. Our testing shows that this design reduces the average measurement time per sample by a factor of 4.5 compared with the pre-optimization configuration.

**G. Integrated control software**

To achieve fully integrated multi-parameter control, we developed a LabVIEW-based software package. Controllable parameters include magnetic field, temperature, sample position, beamsplitter, light source, detector.

Magnetic field control is implemented via communication with the magnet power supply, while temperature control is managed through the VTI temperature controller, which reads calibrated temperature sensors mounted on the sample holder. Sample switching is driven by a stepper motor coupled to the drive shaft; detector switching is controlled by a closed-loop stepper motor driving the translation stage inside the detector chamber. Light source, beamsplitter, and FTIR acquisition parameters are set by modifying the measurement configuration file (*.xpm) through the control program.

The user interface is organized primarily as a command-line sequence (dashed box in Fig. 9). Each command line specifies the settings for a given parameter. The operator first selects the parameter type (e.g., temperature, magnetic field, sample). If the parameter requires a scan, the start value, end value, and step size are entered, while zero of the step size indicates that no scan is needed. Each parameter change is accompanied by a specified ramp rate, wait time, and stability tolerance ("Resolution" in figure). The program adjusts the parameter to the target value at the requested rate and waits for the specified time. If the parameter remains within the tolerance for the entire wait period, it is deemed stable; otherwise, the wait time is reset.

If the "Trigger" option for a given command line is not selected, the program executes the command and proceeds to the next line. If "Trigger" is enabled, the program initiates the FTIR acquisition sequence according to the preselected measurement configuration (e.g., source, beamsplitter) before moving to the next command.

For example, the sequence in Fig. 9 instructs the system to: (1) cool from room temperature to low temperature; (2) switch to sample #2; (3) select the MCT detector; (4) set the globar source; (5) set the KBr beamsplitter; (6) scan the magnetic field from 0 to 12 T in 0.25 T steps, acquiring spectra at each field; (7) switch to the InSb detector; (8) rotate to sample #3; and (9) measure from 12 to 0 T in 0.5 T steps.

Command sequences can be saved as *.csv files and reloaded for repeated experiments. The software continuously displays the current experimental environment and logs key parameters as a function of time, enabling reliable automated execution of complex measurement protocols.

## IV. PERFORMANCE
**A. Signal-to-noise ratio**

The primary objective of the optical throughput enhancement was to improve the signal-to-noise ratio. To quantify the improvement, we measured two magnetic samples, $EuCd_2As_2$ and $EuIn_2As_2$, each with a reflectivity of approximately 40% that exhibits weak field dependence[40]. For each sample, two spectra acquired under identical experimental conditions were divided to yield a noise spectrum ($R_0/R_{0'}$ in Fig. 10). $R_0$ and $R_{0'}$ denote two sequential identical measurements at zero field, respectively.

EuCd$_2$As$_2$ was measured in the Faraday configuration at 8 K and 0 T using a KBr beamsplitter and a liquid-nitrogen-cooled MCT detector (CaF$_2$ beamsplitter and InSb detector for the near-infrared range), with an acquisition time of only 1-2 minutes and a resolution of 4 cm$^{-1}$. EuIn$_2$As$_2$ was measured under similar conditions in the Voigt configuration.

In the absence of noise, the $R_0/R_0$ spectrum should equal unity. Conventional magneto-infrared systems typically exhibit noise levels in the 0.1% to 1% range. In contrast, our optimized system achieves noise on the order of 0.01%. For a 200 cm$^{-1}$ spectral window, the best root-mean-square (RMS) noise averages (red dashed line) 0.01% within 4000-9000 cm$^{-1}$, with a minimum value reaching 0.0061%.

**B. Replica bands in EuCd$_2$As$_2$ (Faraday geometry)**

EuCd$_2$As$_2$ is an antiferromagnetic material at zero field, exhibiting no net magnetization[33,41,42]. Upon applying a magnetic field, spin canting occurs, and the canting angle increases with field strength until, near ~10 T, the system approaches full saturation and becomes ferromagnetically aligned[43]. During this process, exchange interactions cause the electronic structure to evolve with magnetization, leading to corresponding shifts in the energies of interband optical transitions.

Using the near-infrared configuration of our system, we measured the Faraday magneto-infrared spectra of EuCd$_2$As$_2$ with only 1 minute integration time (Fig. 11). A series of optical features were observed to shift with magnetic field in a manner consistent with the magnetization curve, rapid variation at low fields and gradual saturation at high fields. In addition to a dominant high-energy peak, a cascade of weaker replica bands emerged at lower energies. These may originate from spin splitting of the bands or from the Faraday effect.

Such replica bands are typically extremely weak and difficult to detect experimentally. A zoomed-in view shows that their amplitude in $R_B/R_0$ is only about 0.15%, necessitating an RMS noise level on the order of 0.01% to resolve them clearly. The ability to observe these features demonstrates the capability of our optimized setup to detect weak magnetic-field-induced spectral changes.

**C. Landau-level spectroscopy of LaAlSi (Voigt geometry)**

LaAlSi is a topological material in which relativistic quasiparticle excitations may exists[44–46]. When such electronic states lie close to the Fermi level, the application of a strong magnetic field can induce Landau quantization, which can be detected by magneto-infrared spectroscopy. We measured the magneto-infrared response of the (110) surface of LaAlSi in the Voigt geometry using the mid-infrared optical configuration (Fig. 12). The experimental spectra exhibit a series of well-resolved Landau level transitions that shift to higher energies with increasing magnetic field.

These transitions are not linear in field, but share a common nonzero intercept, and the spacing between them decreases as the transition energy increases.

These characteristics are consistent with the Landau quantization of massive Dirac fermions, whose Landau level energies are given by: $E_n = \sqrt{(\Delta^2 + 2n\hbar eBv_F^2)}$, where $\Delta$ is half the energy gap, $n$ is the Landau level index ($n = 0, 1, 2, ...$), $\hbar$ is the reduced Planck constant, $e$ is the elementary charge, $B$ is the magnetic field, and $v_F$ is the Fermi velocity. The corresponding optical transition energies are: $\Delta E_{-n \to n+1} = \sqrt{(\Delta^2 + 2n\hbar eBv_F^2)} + \sqrt{(\Delta^2 + 2(n+1)\hbar eBv_F^2)}$. The excellent agreement between experimental data and this model confirms that the charge carriers along this crystallographic direction form an overall conical band with a finite gap, corresponding to massive Dirac fermion excitations.

## V. CONCLUSION

We have developed a high-flux, high-efficiency magneto-infrared spectroscopy system specifically optimized for broadband measurements in high magnetic fields. Key innovations include precision polishing and gold electroplating of long, narrow light tubes to reduce transmission losses, the adoption of a custom-designed single-on-axis-parabolic-mirror Faraday focusing module to maximize the effective numerical aperture and optical collection efficiency, and the integration of a motorized eight-position sample stage with fully automated control of magnetic field, temperature, optical path, and detector selection. Together, these advances deliver a substantial enhancement in both optical throughput and measurement efficiency, achieving a root-mean-square noise level as low as 0.0061% within 2 minutes for a moderate-reflectivity sample, corresponding to a signal-to-noise ratio exceeding $1.6 \times 10^4$. The system's performance has been validated through the detection of extremely weak magnetic-field-induced spectral features, such as replica bands in $EuCd_2As_2$. These demonstrations highlight the system's capability to probe subtle optical signatures associated with electronic structure evolution and Landau quantization in quantum materials. The design principles presented here can be readily adapted to other high-field optical spectroscopies, offering a versatile platform for advancing precision measurements in condensed matter physics and beyond.

## AUTHOR DECLARATIONS
**Conflict of Interest**
The authors have no conflicts to disclose.

## DATA AVAILABILITY
The data that support the findings of this study are available from the corresponding author upon reasonable request.

**Figure and Tables**

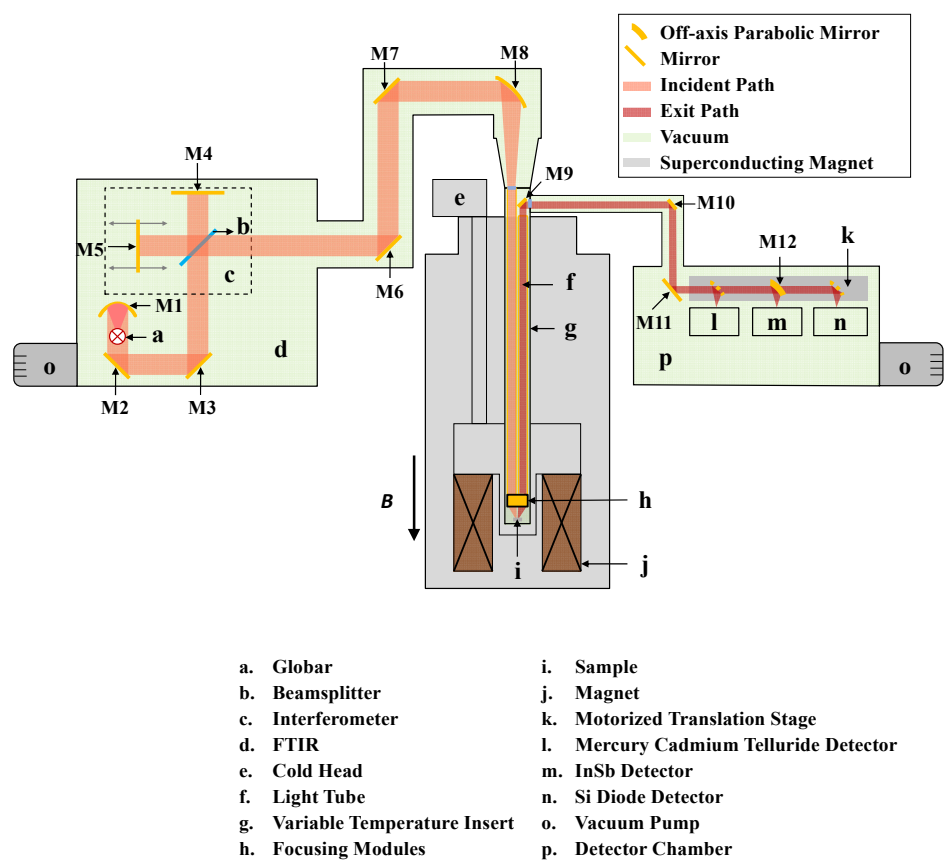

| | |
|---|---|
| a. Globar | i. Sample |
| b. Beamsplitter | j. Magnet |
| c. Interferometer | k. Motorized Translation Stage |
| d. FTIR | l. Mercury Cadmium Telluride Detector |
| e. Cold Head | m. InSb Detector |
| f. Light Tube | n. Si Diode Detector |
| g. Variable Temperature Insert | o. Vacuum Pump |
| h. Focusing Modules | p. Detector Chamber |

**FIG. 1.** Schematic of the high-flux magneto-infrared spectroscopy setup. The system integrates an FTIR spectrometer, a superconducting solenoid, gold-coated light tubes, Faraday and Voigt focusing modules, a motorized eight-position sample stage, and a multi-detector chamber. Infrared radiation sequentially passes through the spectrometer, incident light tubes, focusing modules, sample stage, exit light tubes and external detectors. The light tubes are internally polished and electroplated with gold, reducing the per-meter loss from 65.5% to 22.0%, thereby significantly improving optical throughput.

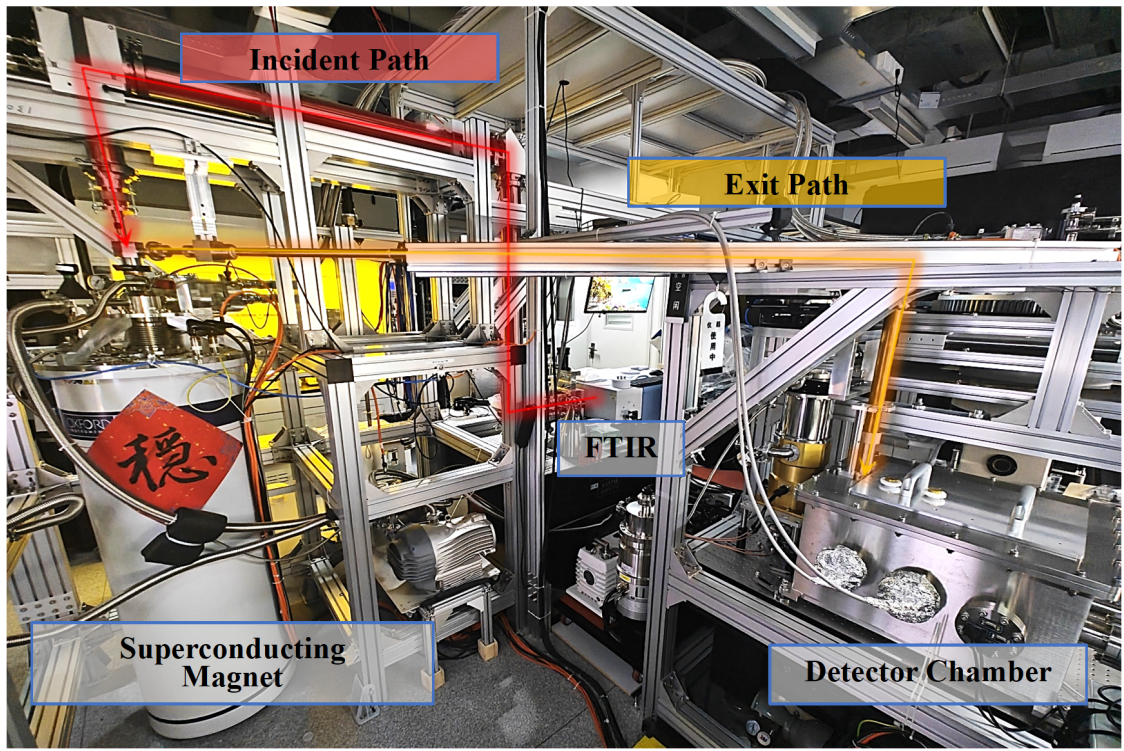

**FIG. 2.** Photograph of the high-flux magneto-infrared spectroscopy system, showing the optical path and the integration of the major subsystems.

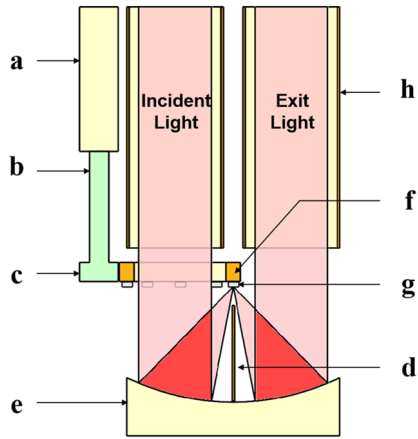
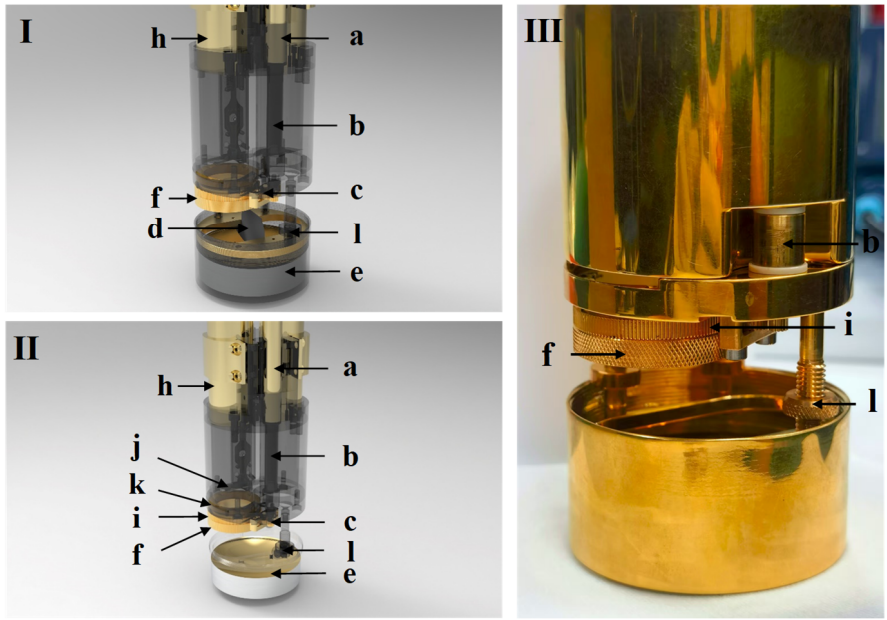

**FIG. 3.** Single-on-axis-parabolic-mirror Faraday focusing module. The incident beam from the light tube is focused onto the sample by an on-axis parabolic mirror, then re-collimated by the same mirror into the exit tube. This design achieves an effective numerical aperture of 0.36 and greatly increases optical collection efficiency within the spatial constraints of the magnet bore. The reflector locking nut is reverse-threaded relative to the mirror tube, so tightening it securely locks the parabolic mirror in place.

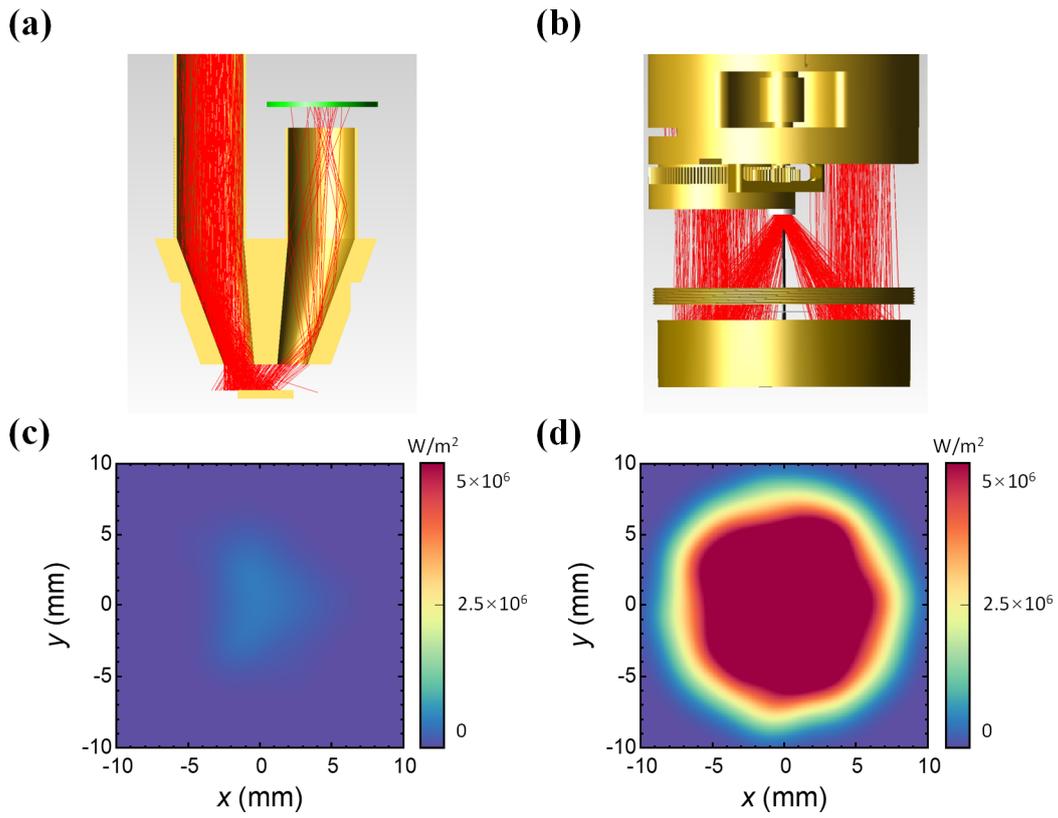

**FIG. 4.** Optical simulation of the single-on-axis-parabolic-mirror focusing design compared with a conventional two-tube scheme. (a,b) Beam propagation simulation. (c,d) Optical power density profile at the exit light tube cross-section. The design improves collection efficiency by nearly an order of magnitude, representing a key element in the system's optical flux enhancement.

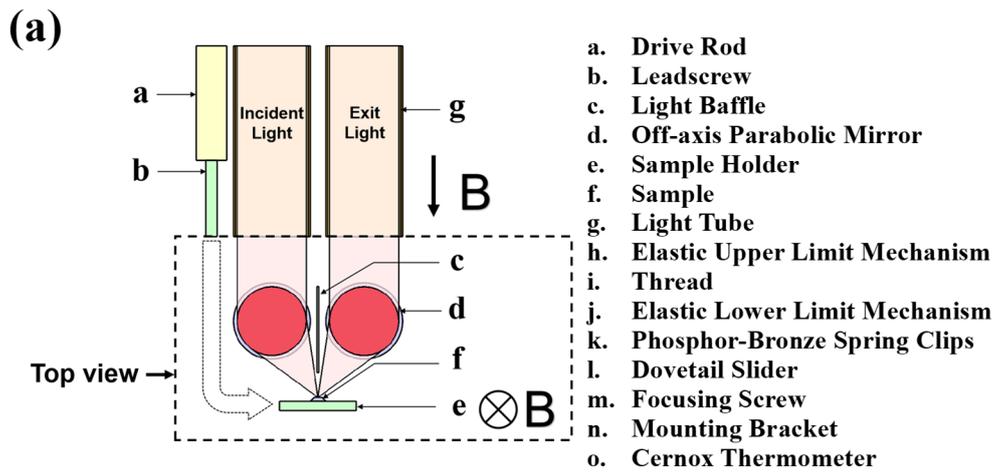

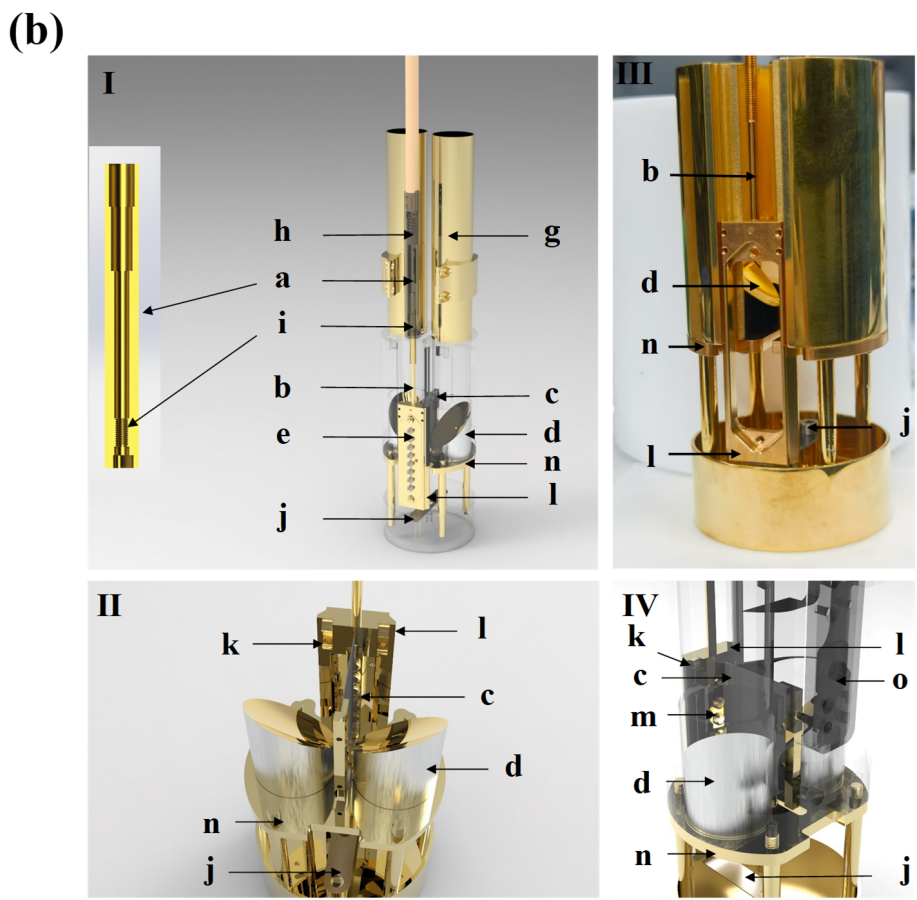

**FIG. 5.** Off-axis-parabolic-mirror Voigt focusing module. The incident beam is directed onto the sample by a 90° off-axis parabolic mirror (OAPM) and collected by a second symmetric OAPM into the exit light tube. The design incorporates a graphite-coated light baffle to suppress stray light and features vertical translation of the sample to sequentially align each position with the shared focal point of the two OAPM.

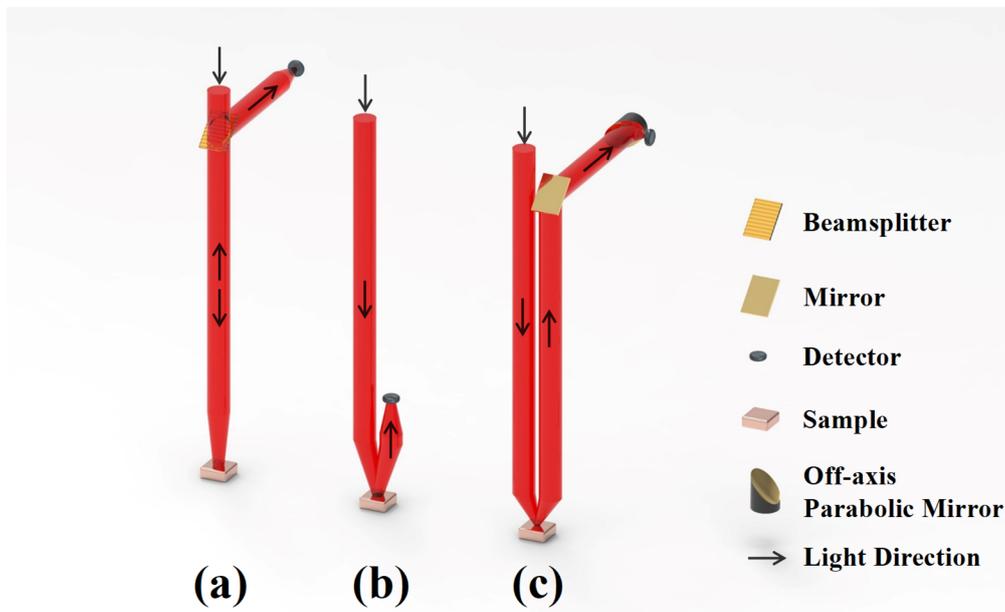

**FIG. 6.** Detector configurations for magneto-infrared spectroscopy. (a) Beam-splitting external-detecting configuration. (b) Reflective internal-detecting configuration. (c) Reflective external-detecting configuration (chosen in this work), which offers high optical flux (avoiding beamsplitter losses), high collection efficiency, and compatibility with both variable-temperature measurements and broadband spectroscopy by allowing detector selection according to the spectral range.

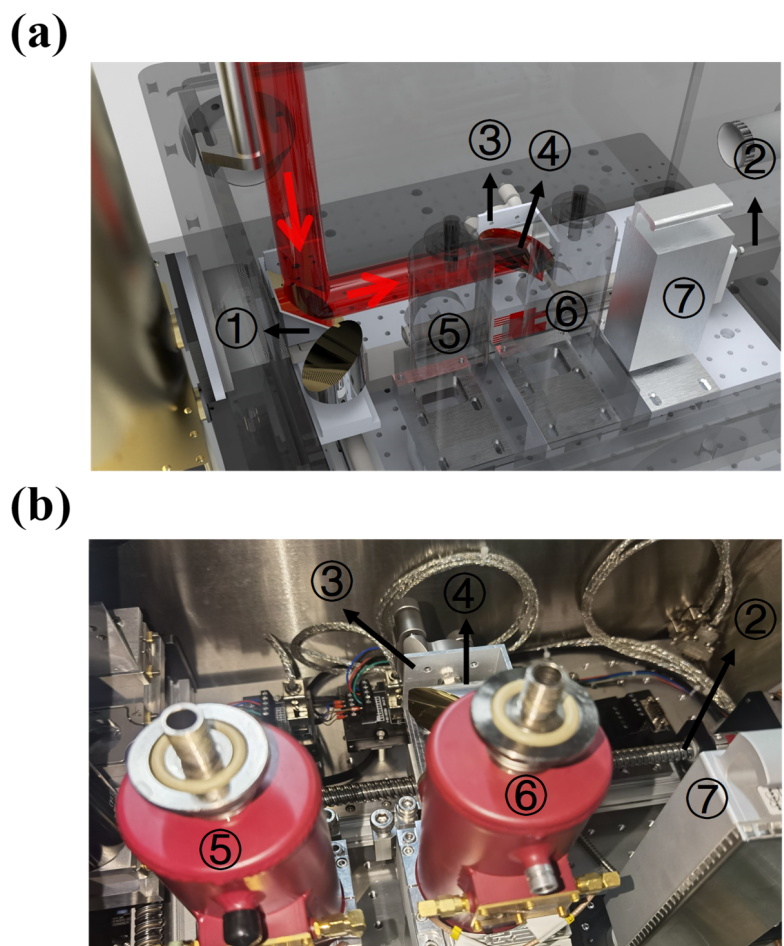

**FIG. 7.** Vacuum-compatible external detector chamber. The collimated output beam is reflected by a large flat mirror and focused by a motorized off-axis parabolic mirror onto the active area of one of three detectors (77 K MCT, 77 K InSb, and Si diode), mounted on a high-precision translation stage. The large effective numerical aperture and in-situ switching improve both collection efficiency and measurement throughput, while maintaining positional stability under vacuum.

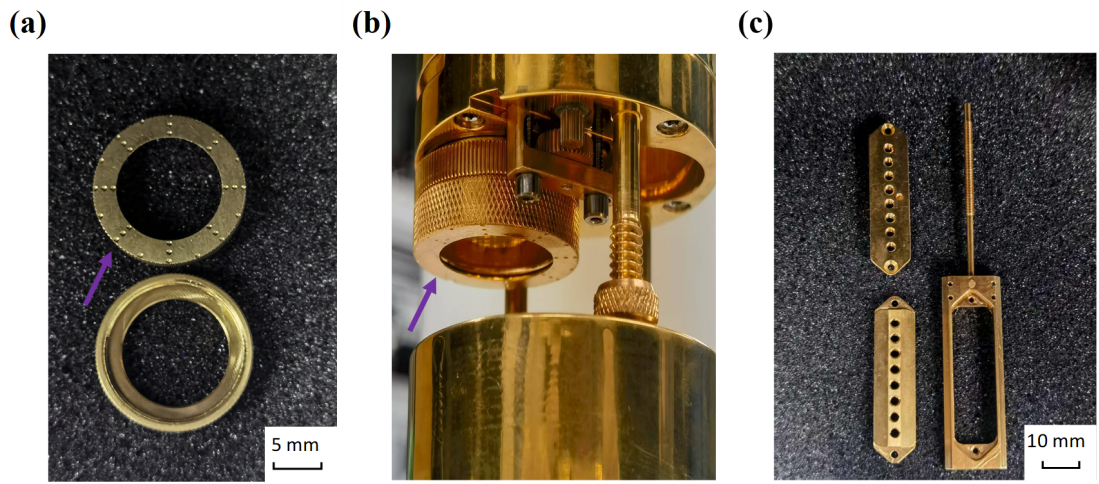

**FIG. 8.** Motorized eight-position sample holders for Faraday and Voigt modules. (a,b) In the Faraday module, a rotating drive rod transmits torque to a gear set that engages the hollow sample ring, sequentially positioning each sample at the focus. (c) In the Voigt module, a translating drive rod moves each sample vertically into the common focal point of the two OAPMs. The modular design accommodates up to eight samples per cooldown, reduces per-sample measurement time by a factor of 4.5, and allows independent preparation and replacement of the sample mount.

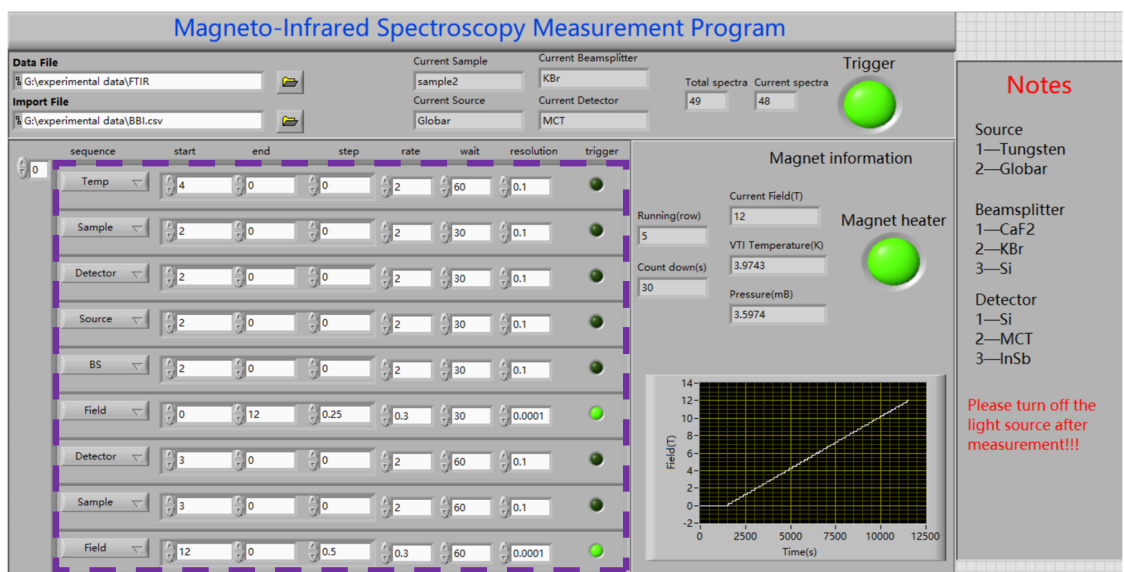

**FIG. 9.** Integrated control software. The LabVIEW-based interface enables automated control of magnetic field, temperature, optical configuration, sample selection, and detector choice, with full synchronization to the FTIR spectrometer for unattended data acquisition.

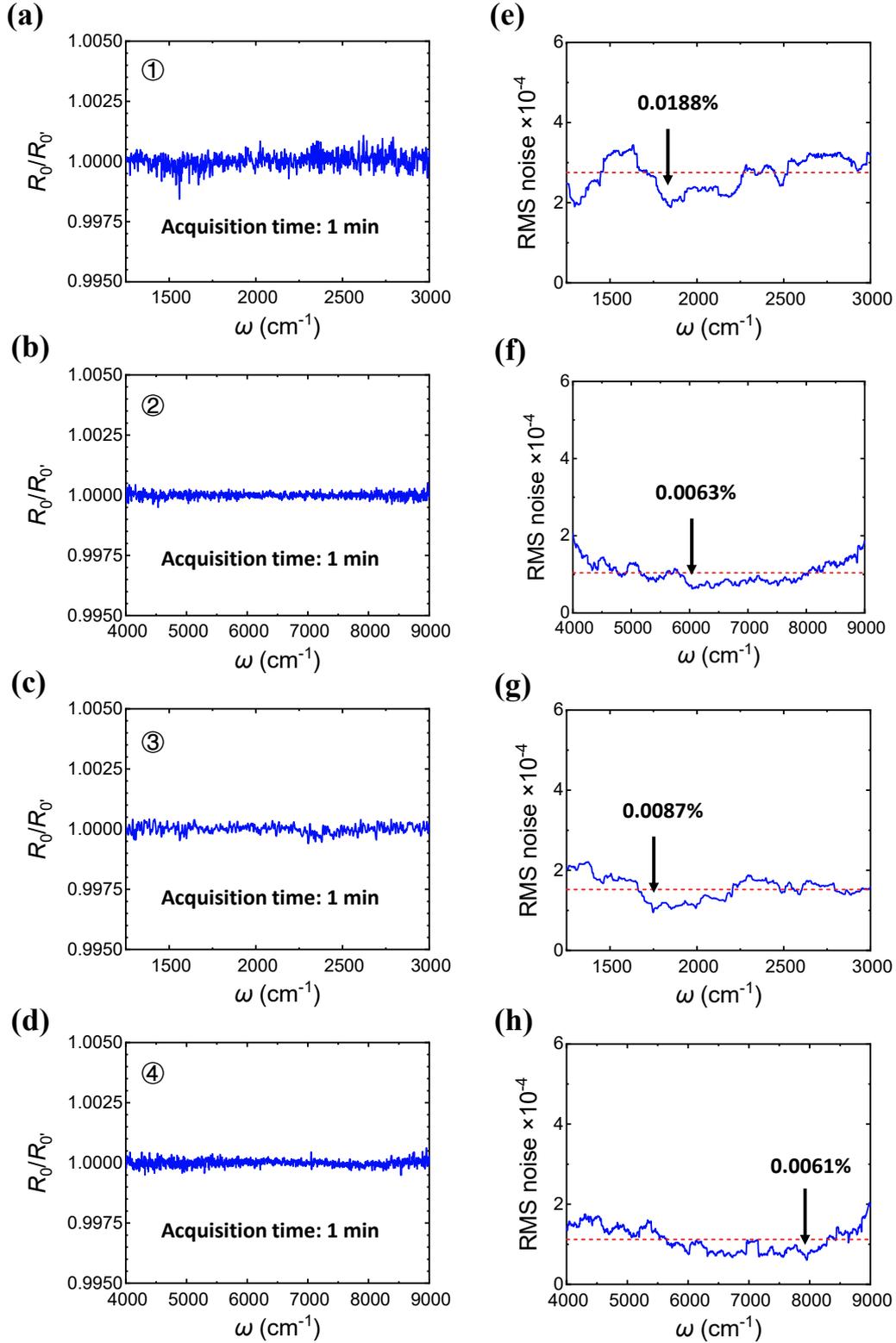

**FIG. 10.** RMS noise level for a moderate-reflectivity sample within a 1-minute acquisition time. (a-d) Noise spectra for the configurations listed in Table 1. The denominator ($R_{0'}$) and numerator ($R_0$) of the vertical-axis label denote the first and second measurements at zero field, respectively. (e-h) RMS noise calculated over a 200 cm$^{-1}$ window. The red dashed line indicates the average RMS noise across each spectrum.

**Table 1 Measurement configuration of RMS noise level test**

| | Sample | Configuration | Temperature | Source | Beam Splitter | Detector | Acquisition Time | Rate | Resolution | RMS Noise Average | RMS Noise Minimum |
|---|---|---|---|---|---|---|---|---|---|---|---|
| ① | $EuCd_2As_2$ | Faraday | 8 K | Globar | KBr | MCT | 1 min | 7.5 kHz | 4 cm$^{-1}$ | $2.75 \times 10^{-4}$ | $1.88 \times 10^{-4}$ |
| ② | $EuCd_2As_2$ | Faraday | 8 K | Tungsten | $CaF_2$ | InSb | 1 min | 7.5 kHz | 4 cm$^{-1}$ | $1.04 \times 10^{-4}$ | $0.63 \times 10^{-4}$ |
| ③ | $EuIn_2As_2$ | Voigt | 5 K | Globar | KBr | MCT | 1 min | 7.5 kHz | 4 cm$^{-1}$ | $1.55 \times 10^{-4}$ | $0.87 \times 10^{-4}$ |
| ④ | $EuIn_2As_2$ | Voigt | 5 K | Tungsten | $CaF_2$ | InSb | 1 min | 7.5 kHz | 4 cm$^{-1}$ | $1.12 \times 10^{-4}$ | $0.61 \times 10^{-4}$ |

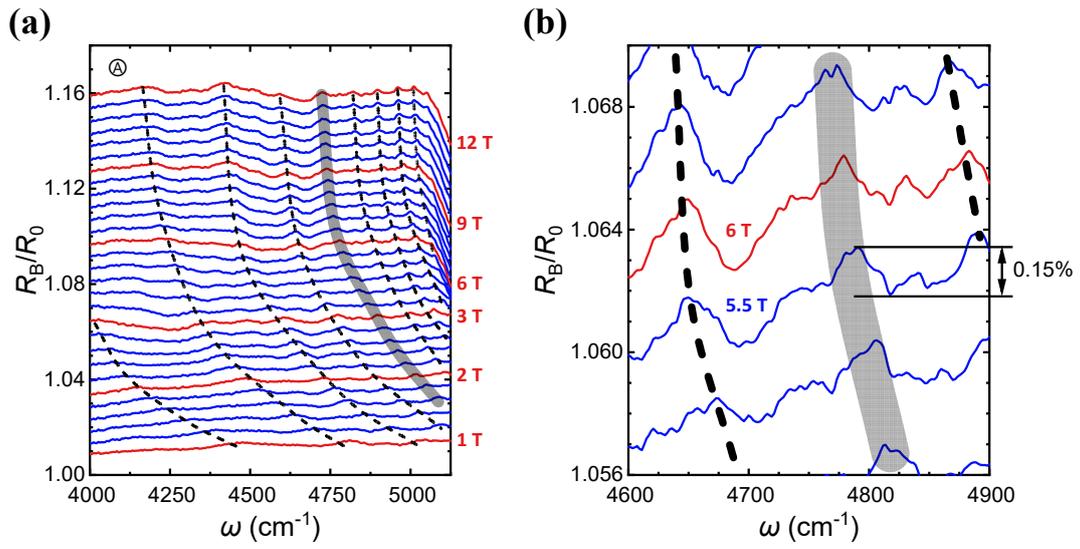

**FIG. 11.** Magneto-infrared spectroscopy of EuCd$_2$As$_2$ in the Faraday geometry. (a) Stacked spectra at different magnetic fields, measured under the configuration in Table 2. (b) Zoom-in view of a weak replica peak with a magneto-optical contrast of 0.15%, clearly resolved and tracked as a function of field, demonstrating the capability of detecting extremely weak field-induced features.

**Table 2 Measurement configuration of EuCd$_2$As$_2$ and LaAlSi**

| | Sample | Configuration | Temperature | Source | Beam Splitter | Detector | Acquisition Time | Rate | Resolution |
|---|---|---|---|---|---|---|---|---|---|
| Ⓐ | EuCd$_2$As$_2$ | Faraday | 8 K | Tungsten | CaF$_2$ | InSb | 1 min | 7.5 kHz | 4 cm$^{-1}$ |
| Ⓑ | LaAlSi | Voigt | 5.7 K | Globar | KBr | MCT | 1 min | 7.5 kHz | 4 cm$^{-1}$ |

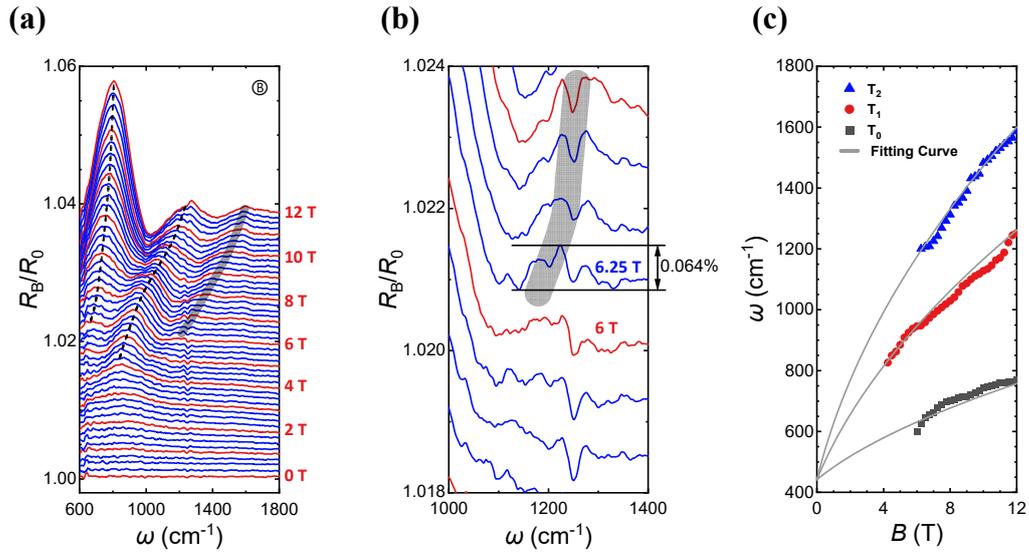

**FIG. 12.** Magneto-infrared spectroscopy of LaAlSi (110) surface in the Voigt geometry. (a) Stacked spectra at different magnetic fields, measured under the configuration in Table 2. (b) Zoom-in view of faint optical features with magneto-infrared amplitude of 0.06%. (c) Extracted Landau-level transition energies compared with the theoretical model for massive Dirac fermions.